# A new theoretical model of optical pumped solid-state laser

Wanfa Liu [1,*]; Yanchao Wang [1]; Shutong He [1]

ABSTRACT

In this paper, from a new perspective of the time (pump time, relaxation time, and stimulated emission time) of the cycle for the active particles at each energy level, the concept of the number of iterative pumping of the laser medium is introduced, and the equivalent model of the laser when it emits light in a steady state is established. By deriving the number of times that the laser medium is repeatedly pumped in the light exit area when the light is emitted, the analytical expression of the laser output power is deduced, and the law of the laser output power changing with the laser parameters is obtained. By fitting and comparing with the experimental results, the new model and the experimental results are in good agreement, the related parameters obtained are also very consistent with the literature, and the changes of multiple parameters and temperature are obtained. These results, especially the expression for the analysis of the laser output power pave a new way for the research and design of optical pumped lasers.

[1] Key Laboratory of Chemical Lasers, Dalian Institute of Chemical Physics, Chinese Academy of Sciences, China

[*] Correspondence to: Key Laboratory of Chemical Lasers, Dalian Institute of Chemical Physics, #457 Zhongshan Road, Dalian, Liaoning Province, China, 116023.
E-mail address: wfliu@dicp.ac.cn (W. Liu).

# A novel theoretical model of optical pumped solid-state laser

ABSTRACT

In this paper, from a new perspective of the time (pump time, relaxation time, and stimulated emission time) of the cycle for the active particles at each energy level, the concept of the number of iterative pumping of the laser medium is introduced, and the equivalent model of the laser when it emits light in a steady state is established. By deriving the number of times that the laser medium is repeatedly pumped in the light exit area when the light is emitted, the analytical expression of the laser output power is deduced, and the law of the laser output power changing with the laser parameters is obtained. By fitting and comparing with the experimental results, the new model and the experimental results are in good agreement, the related parameters obtained are also very consistent with the literature, and the changes of multiple parameters and temperature are obtained. These results, especially the expression for the analysis of the laser output power pave a new way for the research and design of optical pumped lasers.

## I. INTRODUCTION

Optical pumped solid-state lasers due to high efficiencies, compactness and reliability have been widely studied [1-5]. Many kinds of research for the computations and simulations of optic pumped solid-state lasers were developed [6-10]. Rate equations are the main approach to understand temporal variations in the populations of the particles for different energy levels during pumping. As a basic framework of the approach for laser investigating, the results of the computations and simulations about rate equations were consistent with the laser performance on experiments. Thus, the rate equations play an important role in laser researches and design.

However, the unpredictable oscillating modes for the stable resonator make the calculation inconvenient. Moreover, with the pump power increasing, the rising temperature would induce the changing of the absorption coefficient, pump threshold power, and the lifetime of the gain medium. From this point, it is difficult to conform the calculation to the real experiment.

The illustration of the lasing process for the four energy level schemes was shown as Fig. 1. When optical pumped solid-state lasers working at a stable state, within a cycle, the atoms or ions in gain medium would absorb the pump light and were pumped to upper energy level VI (corresponding to the process 1), then down to energy level III by the relaxation process and radiate the laser to make themselves slide into energy level II (corresponding to the process 2, 3), and final, go back to energy level I (corresponding to the process 4). The time of process 1-4 are corresponding to the pump time from energy level I to energy level VI, the relaxation time from energy level VI to energy level III, the stimulated emission time of the gain medium from energy level III to energy level II and the relaxation time from energy level II to energy level I individually. It's interesting to consider the pump time, the relaxation time, and the stimulated emission time of the gain medium. In unit time, the number of cycles is defined as the iterative pump number. With these basic physical parameters, the output power of the laser has been deduced.

In this paper, the dynamic model of lasing was presented. The concept of iterative pump numbers was introduced in this model for the first time. By the derivation for the pump time, the relaxation time, and the total lifetime, the iterative pump number was obtained. The analytic expression for the output power of the diode-pumped solid-state laser was carried out. The output power predicted by this formula is consistent with experimental results. The associated parameters obtained by this model are consistent with previous literature.

## II. EQUIVALENT MODEL FOR STABLE OPERATION LASERS

Diode-pumped Yb: YAG solid-state laser is a typical quasi-four-level system. When the laser is in stable operation condition, the processes of pumping, relaxation, and stimulated emission reach equilibrium, and the numbers of particles at different levels approach constants. That means during the unit time, the number of the particle being pumped to the upper laser level equals the number of stimulated emission photons to the lower laser level, (the relaxation rate from the upper laser level to the lower laser level is several magnitudes slower than the stimulated emission rate and could be ignored here). The number of particles being pumped to the upper laser level during unit time is presumed to be $\Delta n$, and the number of emitted photons is approximately $\Delta n$ as well. If the number of iterative circulation of pump, relaxation, and stimulated emission of laser medium is assumed to be $n$, then during the unit time, the total number of emitted photons is $\Delta n \times n$.

An equivalent model is used in our work: all the laser active particles ($n_0$) were presumed being pumped to the upper laser level, during the unit time; but the total iterative number of the pump, relaxation and stimulated emission is fewer than the model mentioned above, and this number is presumed to be $n'$ ($n'<n$), therefore the total number of the emitted photon $n_0 \times n' \equiv \Delta n \times n$. The iterative number used in this equivalent model is smaller than the real circulation number of laser active particles. This model was successfully used in the prediction of output power and the parameter optimization of diode-pumped solid-state laser. The following section will give the deduction in detail.

### A. Calculation of output power of DPSSL

The three schematics of pumping methods were universal for diode-pumped solid-state lasers as shown in Fig. 2. For the sake of simplicity, we will first treat the case of single-pass end pumping, and then extend the analysis to double-end pumping and double pass end pumping.

Case 1: Single-pass end pumping

The intensity of the pump source is presumed as $I_0$; the absorption coefficient is $\alpha$; the absorption cross-section is $\sigma$; the area of laser medium being pumped by the diode is $S$, $\Delta t$ is the time constant that does not change with the x position.

At position $x$, the inversion photon density $\rho(x)$ to make upper laser level realize population inversion can be expressed as:

$$\rho(x) = I\Delta tS / h\nu_1$$

$$= \frac{(I_0 - I_{th})\Delta tS}{h\nu_1} \exp(-\alpha x - b)/dV \qquad (1)$$

$$= \frac{(I_0 - I_{th})\Delta t}{h\nu_1} \exp(-\alpha x - b)/dx$$

$$\rho(x) = \frac{c_1}{\tau_{1i}(x)\sigma v_V} \qquad (2)$$

$$\tau_1 = \sum \tau_{1i} \qquad (3)$$

$$\tau_1 = \int_0^L \frac{c_1 h\nu_1}{(I_0 - I_{th})\Delta t \sigma v_V} \exp(\alpha x + b) dx \qquad (4)$$

Where $I = (I_0 - I_{th})\exp(-\alpha x - b)$, $b$ is the loss term for pump light as an empirical parameter due to the divergence and diffraction of pump light propagation. This loss term was larger when the laser was pumped by the face diode array and the flash lamp. $v_V$ is the vibrational velocity of the active particles. $\tau_1$ is the total pump time and should be equal to the sum of all the $\tau_{1i}$.

The integral of equation (4) from 0 to $L$ (the thickness of laser active medium) gives:

$$\tau_1 = \frac{h\nu_1}{(I_0 - I_{th})\Delta t \sigma v_V \alpha} [\exp(\alpha L + b) - \exp(b)] \qquad (5)$$

Case 2: Double-end pumping

At position $x$, the inversion photon density $\rho(x)$ can be expressed as:

$$I_{01}(x) = I_0 \exp(-\alpha x - b) \qquad (6)$$

$$I_{02}(x) = I_0 \exp(-\alpha(L - x) - b) \qquad (7)$$

$$I(x) = I_{01}(x) + I_{02}(x) \qquad (8)$$

$$\rho(x) = \frac{(I_0 - I_{th})\Delta t}{h\nu_1} [\exp(-\alpha x - b) + \exp(-\alpha(L - x) - b)]/dx \qquad (9)$$

According to case 1, we obtained $\tau_1$:

$$\tau_1 = \frac{hv_1}{(I_0 - I_{th})\Delta t \alpha \sigma v_V} \exp(\alpha L/2 + b)\left[2\arctan\left(1/\sqrt{\exp(-\alpha L)}\right) - \pi/2\right] \quad (10)$$

Case 3: Double pass end pumping

At position $x$, the inversion photon density can be expressed as:

$$\rho(x) = \frac{(I_0 - I_{th})\Delta t}{hv_1}\left[\exp(-\alpha x - b) + \exp(-\alpha L - b)\exp(-\alpha(L-x) - b)\right]/dx \quad (11)$$

According to case 1, we obtained $\tau_1$:

$$\begin{aligned}\tau_1 = &\frac{hv_1}{(I_0 - I_{th})\Delta t \alpha \sigma v_V} \exp(\alpha L + 3b/2) \\ &\times \left[\arctan(\exp(\alpha L + b/2)) - \arctan(\exp(b/2))\right]\end{aligned} \quad (12)$$

$Yb^{3+}$ doped YAG crystal or ceramic is the quasi-four-level system. The pump time constant $\tau_1$ has been given by formula (5), (10), and (12), and it increases with the decreasing the intensity of pump light. The overall time parameter $\tau$ for a full cycle could be expressed as:

$$\tau = \sum_{i=1}^{4} \tau_i \quad (13)$$

### B. CALCULATION OF THE ITERATIVE PUMP NUMBER

The total iterative pump number $n$ should be inversely proportional to the overall time parameter $\tau$ for a full cycle:

$$n \propto \frac{1}{\tau} = c_1 / \sum_{i=1}^{4} \tau_i \quad (14)$$

where $c_1$ is a period of time and here equals 1s; the sum of $\tau_2$, $\tau_3$ and $\tau_4$ is presumed as $\tau_0$, then equation (14) can be written as:

$$n = \frac{1}{\tau_0 + \tau_1} \quad (15)$$

For three typical pump cases, total iterative pump number $n$ can be written as:

For case 1:

$$n = c_1 \frac{(I_0 - I_{th})\sigma v_V \alpha}{hv_1} \frac{1}{\frac{\exp(\alpha L + b) - \exp(b)}{\Delta t} + \frac{(I_0 - I_{th})\sigma v_V \alpha \tau_0}{hv_1}} \quad (16)$$

For case 2:

$$n = c_1 \frac{(I_0 - I_{th})\sigma v_V \alpha}{h v_1}$$

$$\times \frac{1}{\frac{\exp(\alpha L/2 + b)(2\arctan(\exp(\alpha L/2)) - \pi/2)}{\Delta t} + \frac{(I_0 - I_{th})\sigma v_V \alpha \tau_0}{h v_1}} \quad (17)$$

For case 3:

$$n = c_1 \frac{(I_0 - I_{th})\sigma v_V \alpha}{h v_1} \times$$

$$\frac{1}{\frac{\exp\left(\frac{3\alpha L}{2} + b\right)\left[\arctan\left(\exp\left(\alpha L + \frac{b}{2}\right)\right) - \arctan\left(\exp\left(\frac{b}{2}\right)\right)\right]}{\Delta t} + \frac{(I_0 - I_{th})\sigma v_V \alpha \tau_0}{h v_1}} \quad (18)$$

## C. CALCULATION OF THE UTILITY OF PHOTON

The output power of the laser should always be equal to the total number of active particles $m$ multiplied by the iterative number $n$. For the flowing active particles, the total number of laser particles $m$ is the number of particles in unit time. For the active particles of the solid medium, the iterative number $n$ is the number in unit time.

$$P = h v_2 m n \quad (19)$$

Based on the conservation of energy, the output of the laser can be written as:

For case 1:

$$P = c_1 \frac{v_2}{v_1}(I_0 - I_{th})\sigma v_V \alpha \rho_L SL \frac{1}{\frac{\exp(\alpha L + b) - 1}{\Delta t} + \frac{(I_0 - I_{th})\sigma v_V \alpha \tau_0}{h v_1}} \quad (20)$$

Where $\rho_L$ is the density of active particles.

For case 2:

$$P = c_1 \frac{v_2}{v_1}(I_0 - I_{th})\sigma v_V \alpha \rho_L SL$$

$$\times \frac{1}{\dfrac{\exp(\alpha L/2 + b)(2\arctan(\exp(\alpha L/2)) - \pi/2)}{\Delta t} + \dfrac{(I_0 - I_{th})\sigma v_V \alpha \tau_0}{h v_1}} \quad (21)$$

For case 3:

$$P =$$

$$\frac{c_1 \dfrac{v_2}{v_1}(I_0 - I_{th})\sigma v_V \alpha \rho_L SL}{\dfrac{\exp\left(\dfrac{3\alpha L}{2} + b\right)\left[\arctan\left(\exp\left(\alpha L + \dfrac{b}{2}\right)\right) - \arctan\left(\exp\left(\dfrac{b}{2}\right)\right)\right]}{\Delta t} + \dfrac{(I_0 - I_{th})\sigma v_V \alpha \tau_0}{h v_1}} \quad (22)$$

If supposing $N_0$ satisfies:

$$N_0 = (I_0 - I_{th})\alpha \sigma v_V / h v_1 \quad (23)$$

Due to $\Delta t$ is a constant and can be normalized in the term $\exp(b)$, the formula (20), (21), (22) should be simplified as:

$$P = c_1 h v_2 N_0 \rho_L SL / (\exp(\alpha L + b) - \exp(b) + N_0 \tau_0) \quad (24)$$

$$P = \frac{c_1 h v_2 N_0 \rho_L SL}{\exp(\alpha L/2 + b)(2\arctan(\exp(\alpha L/2)) - \pi/2) + N_0 \tau_0} \quad (25)$$

$$P = \frac{c_1 h v_2 N_0 \rho_L SL}{\exp\left(\alpha L + \dfrac{3}{2}b\right)\left[\arctan\left(\exp\left(\alpha L + \dfrac{b}{2}\right)\right) - \arctan\left(\exp\left(\dfrac{b}{2}\right)\right)\right] + N_0 \tau_0} \quad (26)$$

The formula (24), (25), (26) are universal. In essence, these formulas are the expression of the total emitted photons. If supposing:

$$m = \rho_L SL \quad (27)$$

$$n = \frac{N_0}{\dfrac{\exp(\alpha L + b) - \exp(b)}{\Delta t} + N_0 \tau_0} \quad (28)$$

the formula (20) should be rewritten as:

$$P = c_1 h v_2 m n \eta \quad (29)$$

For the flowing active particles, the total number of laser particles *m* is the number of particles in unit time and *n* is the iterative number in the zone of the gain medium for the laser outputting. For the active particles of the solid medium, *m* is the number of particles in the zone of the gain medium for the laser outputting and the iterative number *n* is the number in unit time. If the laser operating on the optimal condition ($\eta \approx 1$), the output coupling was the optimum.

## III. RESULTS AND DISCUSSIONS

### A. COMPARISON OF THEORETICAL AND EXPERIMENTAL RESULTS

Considering the temperature effect to gain medium, the formula should be rewritten as:

$$P(T) = c_1 \frac{v_2}{v_1}(I_0 - I_{th}(T))\Delta t \sigma^2(T) v_V(T) \rho_L^2 SL$$

$$\times \frac{1}{\frac{\exp(\sigma(T)\rho_L L + b(T)) - 1}{\Delta t} + \frac{(I_0 - I_{th}(T))\sigma^2(T)v_V(T)\tau_0(T)\rho_L}{hv_1}} \quad (30)$$

Where $I_{th}(T) = a_1 + a_2 T$; $b(T) = a_6 + a_7 T$.

The relationship between the absorption cross-section of Yb: YAG crystal and temperature can be described as [11]:

$$\sigma(T) = \left(2.07 + 6.37 \exp\left(-\frac{(T-273)}{288}\right)\right)10^{21} \quad (31)$$

When the temperature of the gain medium increases or decreases, the lifetime $\tau_0$ and the vibrational velocity of the active particles should be a function of temperature. The function could be approximated by Taylor's series. Considering a simple situation, we expanded the function as 1st order Taylor's series and 2nd order Taylor's series:

$$\tau_0(T) = a_8 + a_9 T \quad (32)$$

$$v_V(T) = a_3 + a_4 T + a_5 T^2 \quad (33)$$

### B. THE PARAMETERS OF LASER VARYING WITH THE TEMPERATURE

The parameters of $\tau_0$, $v_V$, $b$ and $I_{th}$ were fitted with the experimental data of the relationships between the laser output power and the pump light power at each temperature point. The experimental data are quoted from [12]. The experiment setup of the literature was a double pass end pumping Yb: YAG laser, the temperature of the gain medium was controlled from 78 K to 300 K and the laser performance has been investigated. Adopting the formula (32) to fitting the curve for the output power, the parameters at each temperature were obtained. Then, the relation of the parameters varying with the temperature was carried out by fitting the parameters at every temperature.

The lifetime $\tau_0$ and vibrational velocity of the active particles in the Yb: YAG gain medium is known to be severely affected by temperature. The lifetime $\tau_0$ and vibrational velocity at every temperature point is obtained by fitting equation (32) and (33). Fig. 3 and Fig. 4 show the lifetime $\tau_0$ for Yb: YAG crystal at 1.03 μm and the vibrational velocity of active particles as a function of temperature. The points are then calculated values and the line is a linear fit. As the temperature decreased from 300 K to 80 K, the lifetime $\tau_0$ for Yb: YAG crystal decreased from 435 μs to 397 μs. The measurement for the luminescent lifetime for Yb: YAG was reported by [13-16]. The value of the luminescent lifetime for Yb: YAG was about two times than the lifetime $\tau_0$ obtained in our model. In fact, the lifetime $\tau_0$ presented in our model contains not only the two of the relaxation time from energy level II to energy level I and from energy level IV to energy level III, but also the stimulated emission time of the gain medium from energy level III to energy level II. The two of the relaxation time usually were shorter and could be ignored. In comparison with no lasing state, the stimulated emission time of the gain medium was less in lasing state. Due to the relations with the velocity of the active particles, the accurate value to obtain was impossible. Nevertheless, the fitting result was close to the experiment measurement in the literature. Using the formula (30) for fitting experimental data under different temperatures, a series of values of the vibrational velocity of active particles of the solid medium was obtained. Fitting the value of the vibrational velocity of active particles of the solid medium by formula (33), the coefficient $a_3$, $a_4$ and $a_5$ could be determined. The function relationship of the lifetime $\tau_0$ and vibrational velocity of active particles of Yb: YAG crystal varying with temperature was linearity. We have not a better explanation about the velocity of the active particles but understanding as the vibration velocity of the active particles, or can only be understood as the vibration velocity of the active particles at present.

Fig. 5 shows the pump light losses as a function of temperature. The points are then calculated values and the line is a linear fit correspondingly. When the temperature decreased from 300 K to 80 K, the pump light losses decreased. The pump light losses varying with temperature was linearity. The threshold pump power density is depicted in Fig. 6. The threshold pump power density at 300 K was found to be approximately 10 kW/cm$^2$. When we adjusted the temperature from 300 K to 70 K, the threshold power at 1030 nm displayed a decreased tendency. The fluctuation is relatively large at low temperatures. This result precisely explains the high efficiency of Yb: YAG at low temperature, and the low threshold design should be selected as much as possible laser design in the future.

For the relationship between the above four parameters and the temperature, the calculation result of our model was consistent with other models and experiment results. Meanwhile, some discrepancy point between the theory and experiments was also displayed from Fig. 3 to Fig. 6. These discrepancies may come from the experiment error and the process of experimental data. The fitted results indicate the quasi-four energy level structure converts to four energy structure, with the temperature decreasing and the discrepancy nearly critical converting zone between the quasi-four energy level structure and four energy structure was obvious.

Furthermore, the next step work was to investigate the correlation of the result with some variables. For instance, the relation between the threshold power density with the temperature and Yb$^{3+}$ doped concentration was out of the forecasting field by our model. This will be further investigated to explain. Meanwhile, this relation had been evidenced by the rate equation [17].A comparative study on pulse pumped mode operation of Yb: YAG laser from cryogenic to room temperatures was presented. Optical to optical efficiency of Yb: YAG lasers as a function of the temperature of Yb:

YAG crystals were depicted in Fig. 7. The solid line shows the calculated results and the dot represents the experiment data. As displayed by Fig. 7, with the temperature falling the optical to optical efficiency was increased. Whereas the decreasing rate was nonlinearity and with the temperature falling to cryogenics the decrease was slower. As indicated by Fig. 7, experimental results have shown a good agreement with the theoretical predictions.

## IV. CONCLUSION

This novel laser model has an ingenious conception. The calculated results and experiments indicated that the computation of this model was coincident with the experimental results. The analytical solution not only could calculate the output power of the laser and the characteristic of laser parameters varying with the experimental condition, but also investigate the characteristic of interior parameters for the laser mechanism by combining the experiment. These characteristics play an important role in the research for the mechanism and technology of lasers.

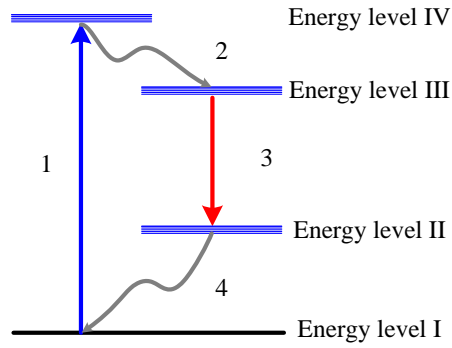

Fig. 1: The illustration of the lasing process for the four energy level schemes.

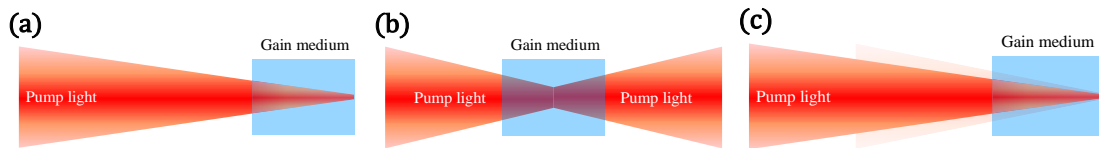

Fig. 2: The three schematics of pumping methods for diode-pumped solid-state lasers: (a) single-pass end pumping; (b) double-end pumping; (c) double pass end pumping.

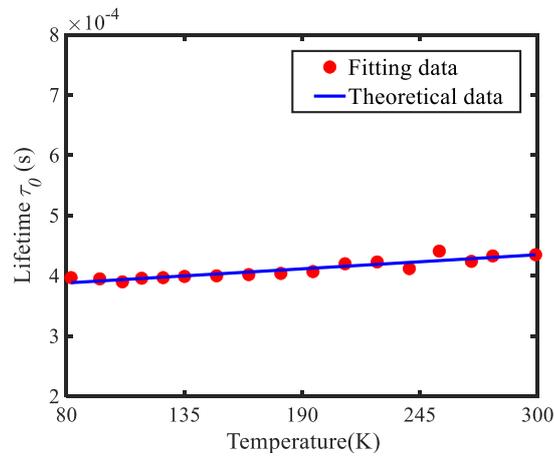

Fig. 3: The lifetime $\tau_0$ for Yb: YAG crystal at 1.03 μm as a function of temperature.

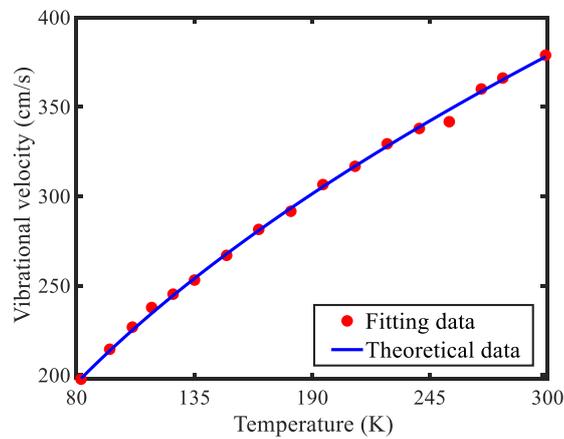

Fig. 4: The vibrational velocity of the active particles of the Yb: YAG crystals as a function of temperature.

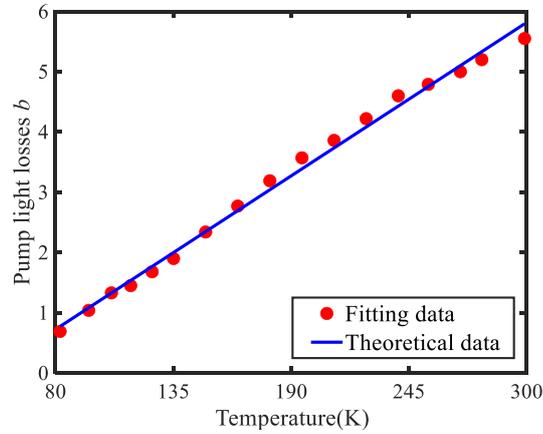

Fig. 5: The pump light losses *b* as a function of temperature.

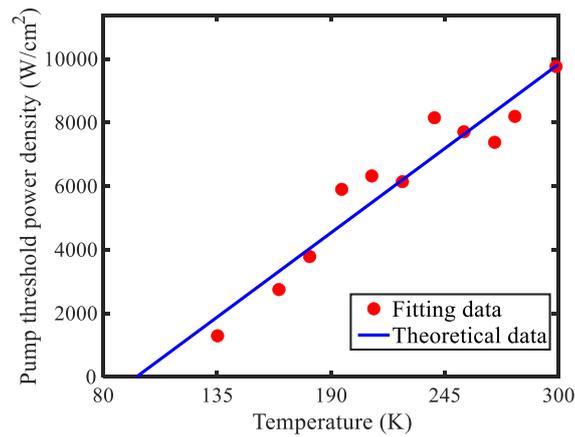

Fig. 6: The threshold pump power density as a function of the temperature.

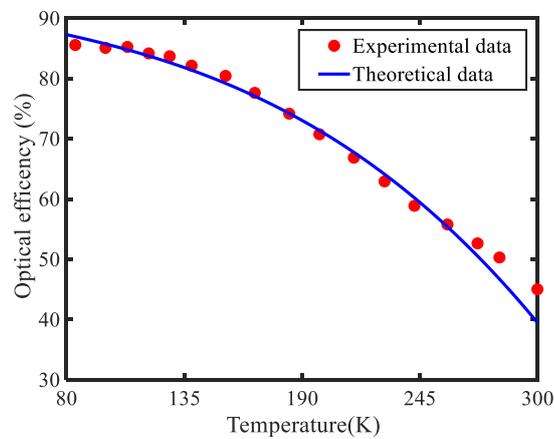

Fig. 7: Optical to optical efficiency of Yb: YAG lasers as a function of the temperature

of Yb: YAG crystals.